# I've Seen Enough: Measuring the Toll of Content Moderation on Mental Health

Gabrielle M. Gauthier
Eesha Ali
Amna Asim
Sarah Cornell-Maier
Lori A. Zoellner

*Department of Psychology*
*University of Washington, Seattle, WA USA*

**ABSTRACT**
**Human content moderators (CMs) routinely review distressing digital content at scale. Beyond exposure, the work context (e.g., workload, team structure, and support) may shape mental health outcomes. We examined a cross-sectional international CM sample ($N = 166$) and a U.S. prospective CM sample, including a comparison group of data labelers/tech-support workers ($N = 45$) and gold-standard diagnostic interviews. Predictors included workplace factors (e.g., hours/day distressing content, culture), cognitive-affective individual differences, and coping. Across samples, probable diagnoses based on validated clinical cutoffs were elevated (PTSD: 25.9–26.3%; depression: 42.1–48.5%; somatic symptoms: 68.7–89.5%; alcohol misuse: 10.5%-18.3%). In the U.S. sample, CMs had higher interviewer-rated PTSD severity ($d = 1.50$), likelihood of a current mood disorder ($RR = 8.22$), and lifetime major depressive disorder ($RR = 2.15$) compared to data labelers/tech-support workers. Negative automatic thoughts ($b = .39 - .74$), ongoing stress ($b = .27 - .55$), and avoidant coping ($b = .30 - .34$) consistently predicted higher PTSD and depression severity across samples and at 3 month follow-up. Poorer perceived workplace culture was associated with higher depression ($b = -.16 - -.32$). These findings strongly implicate organizational context and related individual response styles, not exposure "dose" alone in shaping risk. We highlight structural and technological interventions such as limits on daily exposure, supportive team culture, interface features to reduce intrusive memories, and training of cognitive restructuring and adaptive coping to support mental health. We also connect implications to adjacent "human-in-the-loop" data work (e.g., AI red-teaming), where similar risks are emerging.**

## INTRODUCTION

Content moderators (CMs) review graphic and distressing digital content on behalf of tech employers to evaluate the acceptability of content for online platforms like search engines and social media sites. Reviewing hundreds or even thousands of images, videos, and texts a day for non-permissible, potentially horrific content exposes humans to an unprecedented level of potentially traumatic content; heretofore unheard of and not yet studied in depth in the scientific literature. The mental health impact of such high and repeated levels of potentially traumatic content has received relatively limited empirical investigation (e.g., Spence et al., 2024) despite the high potential for harm and recent exponential growth of this labor force (e.g., Gray & Suri, 2019). Content moderation has often been viewed as a short-term human fix on automation's "last mile" (Gray & Suri, 2019). Yet, ongoing ambiguities and sociocultural dynamics ensure that humans remain "in the loop," making thoughtful work design essential to reducing harm.

The work of repeatedly viewing graphic images such as childhood sexual abuse, acts of terrorism or torture, and sexual assault can be considered exposure to a potentially traumatic event. This was formally recognized by the inclusion of indirect, work-related exposure in the Diagnostic and Statistical Manual of Mental Disorders, 5th edition, diagnosis of posttraumatic stress disorder (PTSD; American Psychiatric Association, 2013). Specifically, indirect exposure was included as a potential Criterion A traumatic event, defined as "experiencing repeated or extreme exposure to aversive details of the traumatic event(s) (e.g., first responders collecting human remains; police officers repeatedly exposed to details of child abuse)" including via work-related "electronic media, television, movies, or pictures" (p. 272). The inclusion of this indirect type of work-related exposure conceptually makes sense, when considering the role of social fear learning.

In PTSD, the trauma (i.e., an unconditioned stimulus) and the unconditioned fear response become associated with internal or environmental cues, which can themselves trigger intrusive memories, flashbacks, physiological arousal, and avoidance (Foa et al., 1989). In contrast, social fear learning refers to learning that occurs, not through direct experience, but through observation of others experiencing aversive consequences when encountering an unconditioned stimulus (Debiec & Olsson, 2017). Observational transmission of fear is an adaptive response to protect oneself and others from experiencing the same or similar aversive consequences in similar situations (Debiec & Olsson, 2017). Simply, experiencing fear and subsequent avoidance of potentially dangerous stimuli based on the observation of others' negative experiences serves a protective function (e.g., fearing and avoiding a poorly-designed, dangerous intersection after seeing a serious car accident there on television) but can also have detrimental mental health effects. Regardless of direct or more indirect exposure, only a small minority of individuals exposed to potentially traumatic events will develop PTSD (e.g., 4.0%; Kessler et al., 2017).

Jobs such as emergency (911) dispatchers, paramedics, journalists, and content moderators all have the potential to experience this type of indirect trauma exposure. While this type of trauma exposure often represents no direct threat to one's own physical integrity, it similarly recruits the amygdala, a key brain structure implicated in fear learning (e.g., Olsson et al., 2007). Notably, observational or informational transmission of fear is generally less likely to result in chronic psychopathology than direct exposure to physical threat (e.g., Monteiro et al., 2016; Pfefferbaum et al., 2019) but nevertheless is still associated with PTSD, depression, somatic symptoms, and substance use problems (e.g., Dadouch & Lilly, 2021; MacDonald et al., 2021; 2023; Monteiro et al. 2016). Although exposure to potentially distressing content is a necessary substrate of moderation work, organizational conditions (e.g., workload pacing, team supports, autonomy, and managerial practices) strongly shape mental health risk—often more than "dose" of exposure alone. This aligns with occupational health models that link job demand/control and organizational justice to depression and PTSD risk (e.g., Theorell et al., 2015; Harvey et al., 2017).

The mental health of content moderators falls within gaps in our scientific knowledge to date. While content moderators experience observational fear learning like other similar occupations, the enormous daily and weekly volume (e.g., hundreds, thousands) of content and the horrific nature of the vast majority of images or videos reviewed stands out as a unique feature of this work. Limited work has examined the incidence of psychiatric disorders, in particular PTSD, depression, alcohol and other substance use in content moderators. In a recent cross-sectional online survey of content moderator mental health ($N = 167$), 53% of those self-identifying as content moderators reported a clinical level of general distress (Spence et al., 2024); with findings replicated in another study by the same researchers, this time specifically recruiting from a technology company pool of content moderators ($N = 160$; Spence & DeMarco, 2025). These surveys highlight the potential detrimental mental health impact of content moderation; however, these studies only screened for general mental health problems (10 items) and wellbeing (7 items) and did not use specific or commonly used psychiatric diagnostic self-report or interview measures. Across these studies, they found a general dose-dependent relationship between potentially traumatic exposure and mental distress, with emotion-focused coping leading to better outcomes and avoidant-focused coping leading to worse outcomes (Spence et al. 2024; Spence & DeMarco, 2025).

Notably, *structural conditions,* including outsourcing, quotas, lack of support, and organizational culture, are likely the primary levers for intervention, not just the nature of the content itself. These factors, some more specific for content moderators, may either increase or decrease the likelihood for developing work-related, content moderation mental health problems. This includes workplace characteristics, cognitive-affect related individual differences, as well as ongoing life stressors and active coping behaviors. With respect to workplace characteristics, "dose" of exposure to explicit content may be important, where a larger dose of work-related trauma exposure often confers greater risk for psychopathology (e.g., Coenen & van der Molen, 2021). Conversely, the supportiveness of the work culture may buffer against mental health problems (e.g., Harvey et al., 2017; Cogan et al., 2024). Further, within the category of work-related experiences, characteristics of daily work-related intrusive memories (for content moderators, intrusive, vivid memories of graphic content) such as their frequency and related distress may predict future mental health problems (e.g., Marks et al., 2018; Michael et al., 2005).

Among psychological trait individual differences, the tendency to use mental imagery, that is, the ability to represent sensory details in the absence of external stimuli in one's mind, has been associated with higher intrusive memories, negative emotions, and PTSD, depression, and anxiety symptoms (e.g., Holmes & Mathews, 2010). Trait empathy while watching someone else go through a distressing scenario is also a risk factor for fear-based pathology (e.g., Debiec et al., 2017; Keum & Shin, 2019; Olsson et al., 2016). Indeed, experimental work manipulating state levels of empathy during social fear learning tasks has shown that empathy moderated the transmission of fear (e.g., Kleberg et al., 2015). Individuals may also differ in their attributional style. A maladaptive appraisal style or dysfunctional attributions reflect how an individual views themselves, the world, and their future (e.g., "I am a failure." "I feel like I am up against the world." "My future is bleak."). The higher presence of this type of cognitive style is consistently a strong predictor of later PTSD (Gómez de La Cuesta et al. 2019) and depression (Fu et al., 2021). Taken together, these individual, potentially trait level differences may put an individual doing content moderation at higher risk for PTSD or depression.

Finally, higher ongoing perceived stress, lower social support, as well as use of maladaptive coping strategies, may also increase the risk for PTSD or depression. Both higher perceived ongoing stress (e.g., Hammen et al., 2009; Vershek-Schallhorn et al., 2015) and lack of social support

(e.g., Brewin et al., 2000; Wang et al., 2018) consistently emerge as predictors of PTSD following trauma exposure and of the onset or relapse of depression. Coping styles, referring to how people handle stress or adversity, may also impact mental health, with adaptive coping strategies such as problem solving and seeking support mitigating the development of long-term mental health problems and maladaptive coping strategies such as avoidance, distraction, and substance use increasing their likelihood (e.g., Kirby et al., 2011; Pineles et al., 2011; Thompson et al., 2010; 2018). Further, one's faith and the use of positive religious coping behaviors such as turning to prayer or finding meaning in difficult situations (e.g., Bjorck et al. 2007; Mahamid & Bdier, 2021) may similarly translate into reduced risk for longer-term mental health problems.

Taken together, the job of content moderation, with its frequent and high content exposure to graphic videos and images, may have detrimental mental health effects, despite not having direct physical threat to the individual performing the job. The present studies sought to use state-of-the-science assessments, including gold standard clinical diagnostic interviews, to better understand PTSD, depression, somatic symptoms, and substance use in this sample. We also examined the association between workplace-related experiences (content moderation workload, frequency and distress associated with work-related intrusive memories, disclosure to others, workplace culture), individual trait differences (use of mental imagery, empathy, and automatic negative cognitions), and ongoing stress and coping (perceived stress, social support, coping strategies, and daily spiritual experiences) on severity of PTSD and depression. The goal of the studies was to better characterize the mental health of content moderators and examine potential predictors of this mental health to aid in promoting resilience and mental health, including designing better computer interfaces supporting moderation and preventing longer-term mental health impacts. In Study 1, given that the vast majority of content moderation is outsourced outside of the U.S., content moderators working for an international contractor ($N$ = 166) completed measures of mental health and also factors potentially cross-sectionally associated with mental health. In Study 2 ($N$ = 45), U.S.-based content moderators working for a major technology company or U.S.-based contractors, and a control sample of technology support workers (e.g., data labelers, customer service), who did not have the expectation that they would process high volumes of graphic content on a daily basis, were compared both on self-report measures but also on clinical diagnostic interviews by trained evaluators. Participants completed follow-up assessments three months later, allowing for both cross-sectional and prospective prediction of PTSD and depression severity. Given concerns about the mental health impact of content moderation, in Study 1 and 2, we sought to characterize clinical levels of PTSD, depression, somatic symptoms, and substance use based on established diagnostic cut-offs. In Study 2, we hypothesized that content moderators would have higher levels of PTSD, depression, somatic symptoms, and substance use than technology support workers, both based on self-report and clinical diagnostic interviews. Due to the limited research to date on content moderators, we did not have *a priori* hypotheses regarding which of the specific predictors of mental health would emerge across samples.

## STUDY 1: INTERNATIONAL CONTENT MODERATOR SAMPLE ($N$ = 166)

### Study 1 Method
*Participants*

Participants were 166 individuals who reported being employed as a content moderator, specifically reviewing distressing, graphic images, videos, or text. The majority (72.3%) self-identified as cisgender male and the overall age of the sample was 27.7 years ($SD$ = 6.27). Participants were recruited through targeted emails distributed to international content moderator contractors affiliated with a major technology company, targeting workers in India, Colombia, Mexico, and the Philippines. The study was described as asking questions about participants' work, beliefs about oneself, and mental health (e.g., anxiety, depression, stress).

Participants must have been between the ages of 18-70 years old, English-speaking, and have worked at least six hours/week as a content moderator for at least the last three months. The definition of content moderator was as follows, "Part of your job is to sort through content that has been flagged for potential removal from a product. While performing this work, you sometimes see, hear, or read content in the form of text, photo, audio, or video files that is potentially distressing, disturbing, upsetting or impactful. You make judgments about whether content must be removed from a product. You have the expectation that some content you encounter may be potentially distressing. This kind of work makes up at least 6 of your work hours in a typical week. You have been doing this kind of work for at least three months."

Individuals who failed two or more attention checks embedded within the survey (e.g. "Select 2 for this item.") were removed from data analysis ($n$ = 15). The final sample size was 166, with 27.7% identifying as cisgender female and 72.3% identifying as cisgender male. On average, participants were 27.7 years of age ($SD$ = 6.3). In terms of racial and ethnic identification, 68.7% ($n$ = 114) identified as South Asian, 14.5% ($n$ = 24) as Pacific Islander, and 9.6% ($n$ = 16) identified as Latino/a. The vast majority (80.7%, $n$ = 134) had completed a bachelor's degree or above.

### Mental Health Outcome Measures
**Posttraumatic Diagnostic Scale for DSM–5 (PDS-5).** This 24-item self-report measure (Foa et al., 2016) assesses lifetime trauma exposure and symptom severity for the past two weeks. DSM-5 Criterion A trauma exposure is assessed with multiple yes/no questions. Each severity item is scored on a Likert scale, ranging from 0 to 4, with higher scores indicating greater symptom severity. A clinical cut score of 27.5 or higher was used as a proxy for a probable PTSD diagnosis (Foa et al., 2016). The PDS-5 has good convergent validity with other self-report and interview measures, as well as high test-retest reliability (Foa et al., 2016).

**Patient Health Questionnaire-9 (PHQ-9).** The PHQ-9 (Kroenke et al., 2001) is comprised of nine items assessing depression. Items are assessed over the past two weeks, rating them on a scale ranging from 0 "*not present*" to 3 "*present nearly every day*." The total score ranges from 0 to 27, with higher scores indicating greater severity of depression. A clinical cut score of 10 or higher was used to indicate moderate or higher depression and probable presence of major depressive disorder (MDD; e.g., Manea et al., 2012). The scale has good convergent validity with interviewer-rated depression severity.

**Somatic Symptom Scale-8 (SSS-8).** The SSS-8 (Gierk et al., 2014) assesses somatic symptom burden (e.g., headaches, stomach, or bowel problems) in the past week, measured across eight items. Each item has a 5-point response scale of 0 to 4, with total scores ranging from 0 to 32. A clinical cut score of 8 or higher was used to capture moderate or higher somatic symptoms (Gierk et al., 2014). The measure has shown distinct and overlapping construct validity with measures of anxiety and depression (Kliem et al., 2021).

**Alcohol Screening Questionnaire (AUDIT-10).** The AUDIT-10 (Saunders et al., 1987, 1993) consists of 10 items assessing the severity of alcohol misuse, focusing on the last two weeks. Each item is scored from 0 "*never*" to 4 "*daily*," with responses ranging from 0 (for non-drinking participants) to a maximum of 40, with a score of 8 or more indicating probable hazardous or harmful alcohol consumption (e.g., Kuitunen-Paul & Roerecke, 2018). The AUDIT has shown good construct validity (Quintero et al., 2019), and excellent test-retest validity (Källmén et al., 2019).

**Cannabis Use Disorders Identification Test-Revised (CUDIT-R).** The CUDIT-R (Adamson et al., 2010) examines the usage and severity of cannabis use over the last two weeks. The present survey used the first eight items, rated on a 0 "*never*" to 4 "*daily or almost daily*" rating. The total scores ranged from 0-32, with higher scores indicating potentially hazardous cannabis use, with a score of nine or higher used to indicate probable clinical levels of cannabis use disorder (e.g., Coelho et al., 2024). The CUDIT-R shows good internal consistency and convergent validity with the other measures of problematic cannabis use and cannabis use disorder (Schultz et al., 2019).

*Workplace-Related Experiences and Culture Measures*

**Workplace-Related Experiences.** The workplace characteristic questionnaire included items assessing features of content moderation duties and workplace supports. The questionnaire included items assessing daily or weekly workload ("*How many hours do you work in a typical week?*") and daily content moderation duties ("*On a typical day, how many of your work hours do you spend sorting potentially distressing/upsetting/impactful content?*"). Additionally, the survey assessed whether participants worked for a major technological company or a secondary vendor ("*tech company,*" "*vendor,*" "*neither*") and whether participants worked from home ("*no,*" "*yes,*" "*combination of working from home and in the office*"). Additional items included disclosure to others and the experience of work-related intrusive memories. In regard to disclosure of their job to others, participants were asked how many people a participant had discussed "*at least some aspect*" of their job. In regard to work-related intrusive memories, an operational definition of intrusive memories was provided, describing them as, "*unwanted thoughts or images that come into your mind. They can be prompted by something you're doing, seeing, hearing, or experiencing, or they can come out of nowhere*" and items asked, in the past two weeks, for the frequency (*0 – 100+*) and associated distress (*1: not at all; 7: extreme*) of these intrusions.

**Workplace Culture Questionnaire.** As part of the workplace experiences survey, participants completed a 13-item questionnaire assessing the extent to which participants perceived that wellness is prioritized in their workplace culture. Content of these items was informed by discussions with content moderators and related administrative staff. See Supplement 1 for measure questions. Items were rated on a 7-point Likert scale (*1: Strongly Disagree – 7: Strongly agree).* Example items included, "*Wellness is part of my company's culture,*" "*I have access to mental health resources at work (e.g., wellness coach, therapy services).*" Items were summed to produce a total score ranging from 13 – 91. For the current sample, internal consistency was good (α = .77).

*Trait Individual Difference Measures*

**Spontaneous Use of Imagery Scale (SUIS).** The SUIS (Nelis et al., 2014) is a self-report questionnaire containing 12 items dedicated to assessing the frequency of visual imagery usage in everyday situations, ranging from "*never*" to "*always.*" Scoring for the SUIS yields a total score ranging from 5 to 35, with higher scores suggesting greater ability in imagery visualization. This measure was included as visual imagery may be related to higher likelihood of experiencing intrusive memories following exposure to potentially traumatic content (Marks et al., 2018). The measure shows convergent validity with related measures (Marks, 1973; Sheehan, 1967) and has good test-retest reliability (ICC = .69; Nelis et al., 2014).

**Interpersonal Reactivity Index (IRI).** This measure (Davis, 1980) assessed empathy and interpersonal sensitivity across several subscales of perspective-taking, fantasy, empathic concern, and personal distress. This measure was included as, particularly from an observer perspective, higher trait empathy may be related to threat memories (Olsson et al., 2016). Each of the 28 items is rated on a five-point scale, ranging from 0 to 4, indicating the degree to which it describes the individual. The measure shows good reliability (Raimondi et al., 2023).

**Automatic Thought Questionnaire (ATQ).** The ATQ (Hollon & Kendall, 1980) is a 30-item tool that measures how frequently one experiences automatic negative thoughts. This measures was included given that trait negative appraisal style has been associated with higher trauma-related intrusive memories (Marks et al., 2018). Each item is rated on a scale from 1 "not at all" to 5 "all the time". The total score of ATQ is calculated by summing up all the items. The ATQ has good convergent validity with other measures of negative beliefs and depressive symptoms (Hill et al., 1989).

*Stress, Social Support, Spirituality, and Coping Measures*

**Perceived Stress Scale (PSS).** This 10-item scale (Cohen et al., 1983) assesses the perception of stressfulness across various life situations in the past month. Ongoing, perceived stress is consistently associated with worse mental health, such as PTSD and depression (e.g., Brewin et al., 2000; Cristóbal-Narváez et al., 2020). Each item is rated on a 5-point scale from 0 "*never*" to 4 "*very often*." The total score is obtained by summing all items, with higher scores indicating higher stress levels. The PSS has good convergent validity with state anxiety (Figalová, & Charvát, 2021).

**Social Connectedness Scale (SCS).** The eight-item SCS (Lee & Robbins, 1995) assesses one's perception of connection and belongingness with others. Social connectedness consistently has been shown to buffer against stress and promote mental health (e.g., Wickramaratne et al, 2022). Responses range from 1 "*strongly agree*" to 6 "*strongly disagree,*" with higher total scores indicating a heightened sense of belongingness. This measure has good convergent validity with social cohesion and self-esteem and discriminant validity with state anxiety (Lee & Robbins, 1998).

**Daily Spiritual Experience Scale (DSES).** The 16-item DSES (Underwood & Teresi, 2002) measures connection with the divine and inner harmony, allowing for the substitution of the word "God" with other concepts. Daily spiritual experiences have been linked to psychological well-being (e.g., Ellison & Fan, 2008). Responses are scored from 1 "*never or almost never* " to 6 "*many times a day,*" with higher scores indicating higher spirituality. The scale has good convergent validity with other measures of spiritual and religious experiences (e.g., Underwood, 2011).

**Brief COPE Scale (BCOPE).** The Brief COPE questionnaire (Carver, 1997) is a commonly used tool to evaluate coping strategies during stressful or difficult times. It comprises 28 items, and can be divided into three coping-style subscales (problem-focused, emotion-focused, avoidant coping). Avoidant coping, in particular, such as using alcohol to cope or giving up on trying to cope, has consistently been associated with higher psychological distress (e.g., Littleton et al. 2007). Respondents rate each item from 1 "*minimal engagement*" to 4 "*high engagement.*" The internal consistency for these three subscales is good (e.g., Littleton et al., 2007).

*Procedure*

This study was reviewed and approved by the University of Washington Institutional Review Board (IRB), with signed consent procedures waived for this study. Participants were provided with an informational consent form prior to participation, where they checked a box that they agreed to participate. Participants then completed study surveys and were paid $50 US dollars via gift card for their participation.

*Data Analysis*

Average severity of mental health problems were examined, along with percentage of participants with scores above a clinical threshold. Simultaneous linear regressions were conducted examining three clusters of cross-sectional predictors of PTSD (PDS-5) and depression (PHQ-9) severity, specifically examining work characteristics (Workplace Experiences and Culture), trait individual differences (SUIS, IRI, ATQ), and, finally, stress, social support, and coping (PSS, SCS, DES, BCOPE).

**Study 1 Results: International Content Moderators Sample ($N = 166$)**

*Descriptive Information*

Only 19.3% of international content moderators reported being employed by a major technology company, another 28.9% employed by a vendor, with the majority working for neither (51.8%). Most (85.5%) worked from home or the combination of home and office, and approximately half (54.2%) reported working within a team. The vast majority (78.8%) earned the equivalent of $30,000 USD or less per year. On average, content moderators reported working 32.84 hours/week ($SD = 11.69$), with 3.0 hours/day ($SD = 2.16$) sorting potentially distressing, upsetting images. This included moderating pornographic/explicit content (89.8%, $n = 149$), terrorism-related content (68.1%, $n = 113$), and childhood physical or sexual abuse content (65.7%, $n = 109$).

*Mental Health: PTSD, Depression, Somatic Symptoms, and Substance Use*

On average, content moderators reported experiencing 12.96 ($SD = 15.01$) work-related intrusive memories in the past two weeks, with moderate related distress ($M = 3.81$, $SD = 1.62$; *not at all* [1] to *extremely* [7]) associated with the worst work-related intrusive memory. Means and standard deviations can be seen in Table 1, generally reflecting non-clinical levels of PTSD, depression, somatic symptoms, and problematic substance use. However, based on standardized

**Table 1.** *Mental Health Means and Standard Deviations for International Content Moderator Contractor Sample (N = 166)*

| Measure | Mean | SD | Range | Clinical Diagnostic Level % | n |
|---|---|---|---|---|---|
| PTSD Severity Full (PDS-5) | 14.33 | 18.13 | 0 - 65 | 25.9 | 43 |
| PTSD Severity (PDS-5)[1] Trauma Exposed | 26.27 | 16.99 | 0 - 65 | 47.8 | 43 |
| Depression (PHQ-9) | 9.61 | 5.86 | 0 - 24 | 48.5 | 80 |
| Somatic Symptoms (SSS-8) | 11.17 | 6.20 | 0 - 28 | 68.7 | 114 |
| Alcohol Use (AUDIT-10) | 3.45 | 5.08 | 0 - 27 | 18.3 | 30 |
| Cannabis Use (CUDIT-R) | 0.57 | 2.20 | 0 - 14 | 2.4 | 4 |

*Note.* [1]$n = 90$ with DSM-5 Criterion A Trauma; PDS-5: Posttraumatic Diagnostic Scale for DSM–5 (> = 27.5, clinical cutoff); PHQ-9: Patient Health Questionnaire-9 (> = 10, clinical cutoff); SSS-8: Somatic Symptom Scale-8 (> = 8 moderate to severe); AUDIT-10: Alcohol Screening Questionnaire-10 (> = 8, hazardous use cutoff); CUDIT-R: Cannabis Use Disorders Identification Test-Revised (> = 9, clinical cutoff).

**Table 2.** *Work-Related Experiences, Individual Difference, and Stress Predictors of PTSD Severity (N = 166)*

| Predictors of PTSD Severity[1] | B | SE | Beta | 95% CI LB | 95% CI UB | p |
|---|---|---|---|---|---|---|
| Work Experiences[2] | | | | | | |
| Hours/day Distress Content | 0.64 | 0.64 | .09 | -0.64 | 1.91 | .32 |
| # People Disclose | -1.33 | 0.60 | -.21 | -2.51 | -0.15 | .03* |
| Work-related Intrusions | 0.13 | 0.10 | .13 | -0.07 | 0.32 | .20 |
| Intrusion-related Distress | 4.93 | 0.98 | .49 | 2.98 | 6.88 | <.001* |
| Work Culture | -0.09 | 0.15 | -.06 | -0.39 | 0.21 | .54 |
| Trait Individual Differences[3] | | | | | | |
| Use of Imagery (SUIS) | 0.50 | 0.20 | .22 | 0.10 | 0.90 | .01* |
| Interpersonal Reactivity (IRI) | -0.22 | 0.14 | -.16 | -0.49 | 0.05 | .11 |
| Automatic Thoughts (ATQ) | 0.36 | 0.05 | .67 | 0.27 | 0.46 | <.001* |
| Stress, Coping and Support[4] | | | | | | |
| Daily Perceived Stress (PSS) | 0.68 | 0.25 | .27 | 0.20 | 1.17 | .007* |
| Social Support (SCS) | -0.45 | 0.15 | -.29 | -0.74 | -0.16 | .003* |
| Daily Spiritual Exp (DSES) | -0.04 | 0.08 | -.04 | -0.20 | 0.13 | .66 |
| Problem-focus Coping (BCOPE) | 0.51 | 0.40 | .15 | -0.29 | 1.32 | .21 |
| Emotion-focus Coping (BCOPE) | -0.25 | 0.38 | -.08 | -1.00 | 0.51 | .52 |
| Avoidant Coping (BCOPE) | 1.10 | 0.35 | .31 | 0.41 | 1.79 | .002* |

*Note.* SUIS: Spontaneous Use of Imagery; IRI: Interpersonal Reactivity Index; ATQ: Automatic Thoughts Questionnaire; PSS: Perceived Stress Scale; SCS: Social Connectedness Scale; DSES: Daily Spiritual Experiences; BCOPE: Brief COPE Scale.
[1] DSM-5 Criterion A trauma-exposed sample based on PDS-5 (*n* = 90)
[2] $F(5, 84) = 7.87, p < .001, R^2 = .32$
[3] $F(3, 85) = 22.86, p < .001, R^2 = .45$
[4] $F(6, 83) = 10.61, p < .001, R^2 = .43$

**Table 3.** *Work Experiences, Individual Difference, and Stress Predictors of Depression Severity (N = 166)*

| Predictors of Depression Severity | B | SE | Beta | 95% CI LB | 95% CI UB | p |
|---|---|---|---|---|---|---|
| Work Experiences[1] | | | | | | |
| Hours/day Distressing Content | 0.27 | 0.20 | .10 | -0.12 | 0.65 | .17 |
| Number People Disclose Job | -0.39 | 0.17 | -.17 | -0.72 | -0.06 | .02* |
| Work-related Intrusions | 0.02 | 0.03 | .05 | -0.04 | 0.08 | .52 |
| Intrusion-related Distress | 1.26 | 0.27 | .36 | 0.73 | 1.79 | <.001* |
| Work Culture | -0.09 | 0.04 | -.16 | -0.16 | -0.01 | <.03* |
| Trait Individual Differences[2] | | | | | | |
| Use of Imagery (SUIS) | 0.06 | 0.04 | .08 | -0.02 | 0.14 | .16 |
| Interpersonal Reactivity (IRI) | 0.01 | 0.03 | .02 | -0.04 | 0.06 | .75 |
| Automatic Thoughts (ATQ) | 0.15 | 0.01 | .74 | 0.13 | 0.17 | <.001* |
| Stress, Social Support, Coping[3] | | | | | | |
| Daily Perceived Stress (PSS) | 0.40 | 0.06 | .46 | 0.29 | 0.51 | <.001* |
| Social Support (SCS) | -0.08 | 0.03 | -.16 | -0.15 | -0.02 | .01* |
| Daily Spiritual Exp (DSES) | -0.01 | 0.02 | -.04 | -0.05 | 0.03 | .57 |
| Problem-focus Coping (BCOPE) | -0.10 | 0.09 | -.09 | -0.29 | 0.09 | .29 |
| Emotion-focus Coping (BCOPE) | -0.03 | 0.09 | -.03 | -0.21 | 0.15 | .73 |
| Avoidant Coping (BCOPE) | 0.39 | 0.09 | .30 | 0.22 | 0.56 | <.001* |

*Note.* SUIS: Spontaneous Use of Imagery; IRI: Interpersonal Reactivity Index; ATQ: Automatic Thoughts Questionnaire; PSS: Perceived Stress Scale; SCS: Social Connectedness Scale; DSES: Daily Spiritual Experiences; BCOPE: Brief COPE Scale.
[1] $F(5, 156) = 8.40 \, p < .001, R^2 = .21$
[2] $F(3, 158) = 72.71, p < .001, R^2 = .58$
[3] $F(6, 157) = 24.08, p < .001, R^2 = .48$

clinical cut-offs, approximately half the sample met criteria for likely diagnosis of PTSD (47.8%) and major depressive disorder (48.5%). Further, 68.7% reported moderate to severe somatic symptoms. Eighteen percent (18.0%) reported hazardous alcohol use, while only 2.4% reported problematic cannabis use. Taken together, on average, international content moderators had generally low severity scores on well-established, psychometrically-validated self-report measures of mental health; yet, the percentage of individuals who scored above clinical-cut off levels suggested the presence of potential clinical diagnoses of PTSD, major depression, and somatization disorders (47.8%-68.7%) was noteworthy.

*Cross-sectional Predictors of Mental Health*

**Work-Related Experience Predictors.** When examining work-related experiences as predictors of mental health, as seen in Table 2, for those who had experienced a DSM-5 Criterion A trauma, disclosing details to fewer individuals about one's work ($b$ = -.21, $p$ = .03) and higher distress about the worst work-related intrusion in the past two weeks ($b$ = .49, $p$ < .001) were associated with higher PTSD severity (PDS-5). Similarly, as seen in Table 3, disclosing details to fewer individuals about one's work ($b$ = -.21, $p$ = .03), higher distress about the worst work-related intrusion in the past two weeks ($b$ = .36, $p$ < .001), and worse work-related culture ($b$ = -.16, $p$ = .03) were associated with higher depression severity (PHQ-9).

**Individual Difference Trait Predictors.** When examining individual difference trait predictors of mental health, as seen in Table 2, for those who had experienced a DSM-5 Criterion A trauma, higher use of imagery ($b$ = .22, $p$ = .03) and higher negative automatic thoughts ($b$ = .67, $p$ < .001) were associated with higher PTSD severity (PDS-5). As seen in Table 3, higher negative automatic thoughts ($b$ =

.74, *p* < .001) was strongly associated with higher depression severity (PHQ-9).

**Stress, Social Support, and Coping Predictors.** When examining stress and coping predictors of mental health, as seen in Table 2, for those who had experienced a DSM-5 Criterion A trauma, higher perceived stress in the last month (*b* = .27, *p* = .007), lower social connectedness (*b* = -.29, *p* = .003), and more avoidant coping (*b* = .31, *p* = .002) were associated with higher PTSD severity (PDS-5). As seen in Table 3, higher perceived stress in the last month (*b* = .46, *p* < .001), lower social connectedness (*b* = -.16, *p* = .01), and higher avoidant coping (*b* = .30, *p* < .001) were associated with higher depression severity (PHQ-9).

Taken together, some of the strongest cross-sectional predictors of higher PTSD or depression were distress related to work-related intrusive memories, the presence of higher automatic negative thoughts, and higher perceived stress. Poorer work-related culture, avoidant coping style, and lower social connectedness to others also were associated with higher PTSD or depression severity.

## STUDY 2: U.S.-BASED CONTENT MODERATORS AND TECHNOLOGY SUPPORT WORKERS (*N* = 45)

### Study 2 Method
*Participants*

Participants were 45 individuals who reported current employment as a content moderator or in technology support/data labeling. Participants were recruited through targeted emails distributed to content moderator contractors affiliated with major technology companies or their vendors, postings on work-related social media outlets, and community flyers targeting individuals who work in these positions. Inclusion criteria included current employment as a content moderator or technology support/data labeler. The same definition of content moderator was used as in Study 1. The technology support/data labeler role was defined as, "Part of your job is to sort through content (e.g. text, photos, videos) and apply descriptive labels, or you provide technology related customer service. You do not have any expectation that the content you encounter will be potentially distressing. You do not perform any content moderation work (as described above) and you perform data labeling/ technology support for at least 6 of your work hours in a typical week. You have been doing this kind of work for at least three months." Participants must have been between the ages of 18-70 years old, English-speaking, lived in the greater Seattle area, and worked at least six hours/week as a content moderator or technology support/data labeler for at least the last three months.

On average, content moderators (*n* = 19) had a mean age of 31.37 years (*SD* = 7.71) and technology support/data labelers (*n* = 26) had a mean age of 32.00 years (*SD* = 5.95). Approximately half of the content moderators (47.1%) and technology support/data labelers (50.0%) self-identified as cisgender female. Approximately half (52.6%) of the content moderators and 61.5% of the technology support/data labelers self-identified as from European ancestry. For content moderators, 15.4% identified as from Asian ancestry and 15.8% as of Latino/a ethnicity. For technology support/data labelers, 26.3% identified as from Asian ancestry and 19.2% as of Latino/a ethnicity. The vast majority had completed a bachelor's degree or higher (78.9% content moderators, 80.8% technology support/data labeler).

*Measures*

**Interviewer-Rated Mental Health Assessments.** Semi-structured interviews were conducted by well-trained clinical interviewers, who were receiving doctoral level clinical psychology training. Each interviewer was trained to an 80% reliability criterion on each of the interview measures, using three standardized cases. They further attended weekly clinical supervision, where, for each participant, each measure was presented to a clinical team where licensed clinical psychologists and other trained interviewers reviewed their clinical judgments and ratings. Over the course of the study, routine inter-rater reliability meetings occurred where assessors rated the same case and discussed clinical ratings.

*Posttraumatic Stress Symptom Inventory for DSM-5 (PSS-I-5).* The PSS-I-5 (Foa et al., 2016) 24-item scale is an interview rated tool that assesses PTSD symptom severity in the past two weeks. The brief trauma history screen from the Structured Clinical Interview for DSM-5 (SCID-5; First et al., 2016) was used to ascertain DSM-5 Criterion A trauma exposure. Of note, the DSM-5, includes an A4 criteria explicitly including "experiencing repeated or extreme exposure to aversive details of the traumatic event(s) (e.g., first responders collecting human remains; police officers repeatedly exposed to details of child abuse) (p. 271, DSM-5, 2013)." Of note, "Criterion A4 does not apply to exposure through electronic media, television, movies, or pictures, unless this exposure is work related" (p. 271, DSM-5, 2013). Accordingly, content moderators who repeatedly moderate work-related distressing content could qualify as potentially being exposed to DSM-5 Criterion A trauma. For those who met DSM-5 Criterion A trauma exposure, 20 PTSD symptom items were rated on a 5-point scale for the last two weeks. Content moderation-related potential trauma exposure was specifically queried and related PTSD symptoms were assessed, if present. The PSS-I-5 yields a total severity score, ranging from 0 to 80, and a PTSD diagnosis using the DSM-5 polythetic diagnostic criteria. The PSS-I-5 shows good test re-test reliability and robust convergent validity with other interviewer and self-report measures of PTSD (Foa et al., 2016).

*Quick Inventory of Depressive Symptomatology Interviewer Rating (QIDS-I).* The QIDS-I (Rush et al., 2003) 16-item interview version assesses DSM criterion symptom domains for a major depressive episode, serving as a measure of depression severity. Interviewers rate each item from 0 to 3, with higher scores indicating more severe symptoms. The total score ranges from 0 to 27. The QIDS-I has good convergent validity with other interview and self-report measures of depression severity (Trivedi et al., 2004).

**Structured Clinical Interview for DSM-5 (SCID-5 Research Version).** The SCID-5 (First et al., 2016) is a semi-structured interview that systematically assesses DSM-5 major psychiatric disorders. The SCID-5 was included as a well-validated clinical diagnostic interview measure. Each item or criterion is scored as 0 for "*absent,*" 1 for "*subthreshold,*" or 2 for "*threshold,*" with items combined to determine if a DSM-5 diagnosis was present. In particular, the presence of any mood disorder, major depressive disorder specifically, and substance use disorders were purposely examined. The SCID-5 has good inter-rater reliability, with high positive agreement on the presence of a diagnosis (Osório et al., 2019).

**Self-Report Mental Health Measures.** The same set of self-report measures of mental health (PDS-5, PHQ-9, SSS-8, AUDIT, CUDIT-R), workplace-related experiences, trait individual differences (SUIS, IRI, ATQ), and stress, coping, and support (PSS, SCS, DSES, BCOPE) were administered.

### Procedure

This study was reviewed and approved by the University of Washington Institutional Review Board (IRB), with signed consent procedures. At baseline and three month follow-up, participants met with a clinician trained to reliability in the PSS-I-5 and SCID-5, to assess the presence of potential PTSD and its severity as well as the presence of other DSM-5 diagnoses, both current and lifetime. All PSS-I-5s and SCID-5s were reviewed in a weekly, clinical team supervision meeting, supervised by licensed clinical psychologists. Participants also completed self-report measures described above at both timepoints. Participants received $200 USD for completing the self-report measures and clinical interviews.

### Data Analysis

Mean mental health severity scores were calculated, including both self-report mental health outcomes for parity with Study 1 and percentage endorsed for interviewer-rated gold-standard diagnostic information (SCID-5). Probable clinical diagnosis was calculated for self-report measures based on established clinical cut-offs. Observed clinical diagnosis were reported for PTSD, mood disorders, and substance use disorders. Simultaneous linear regression was performed to assess the cross-sectional and prospective relationship between hypothesized baseline predictors in the full sample and self-reported PTSD and depression severity (PDS-5, PHQ-9). For consistency and for generalizability of results, the same sets of predictors in Study 1 were used in Study 2. However, given the smaller sample size in Study 2, for coping (BCOPE), only avoidant coping but not problem-focused or emotion-focused coping were included in the regression equations.

### Study 2 Results: U.S.-based Content Moderators and Technology Support Workers ($N = 45$)

#### Descriptive Information

For the content moderators ($n = 19$), 42.1% reported being directly employed by a major technology company rather than a vendor. Fifty-eight percent earned the equivalent of $90,000 USD or more per year. Eighty-four percent reported working from home or a combination of home and office, with 94.7% working within a work team. On average, content moderators reported working 37.74 hours/week ($SD = 9.91$), with 4.68 hours/day ($SD = 2.63$) sorting potentially distressing, upsetting images. This included moderating pornographic/explicit content (94.7%, $n = 18$), terrorism related content (78.9%, $n = 15$), and childhood physical or sexual abuse content (89.5%, $n = 17$).

Technology support/data labeler workers ($n = 26$) did not differ from content moderators on the likelihood of being employed by a major technology company (38.5%) rather than a vendor, average income (53.7% annual income above $90,000 USD), working from home or a combination of office and home (84.6%), on a team (92.3%), or hours worked per week ($M = 37.27$, $SD = 9.94$). Nor were they different on age, gender, or education level. As would be expected, the hours spent on monitoring distressing or explicit content and the nature of that content (pornographic, terrorism, childhood abuse) differed (all $p$s < .001).

### Mental Health: PTSD, Depression, Somatic Symptoms and Substance Use

**Posttraumatic Stress and Depression.** While content moderators and tech support workers did not differ in the number of work-related intrusive memories in the past two weeks, they did differ in distress related to the worst intrusion (Content Moderators: $M = 3.00$, $SD = 1.73$; Tech Support: $M = 1.52$, $SD = 1.71$, $F(1, 42) = 7.99$, $p = .007$, $d = 0.86$). As seen in Table 4, for self-reported PTSD severity (PDS-5), content moderators and technology support workers did not differ on severity or likelihood of exceeding a clinical cut score. Content moderators were more likely to have reported work-related content that could potentially meet criteria for a DSM-5 criterion A trauma (100%, $n = 19$) than tech support workers (3.8%, $n = 1$), $\chi^2(N = 45) = 41.11$, $p < .001$, $RR = 13.5$.

**Table 4.** *Mental Health Means and Standard Deviations for U.S. Content Moderator vs Tech Support/Data Labeler Sample ($N = 45$)*

| Measure | Mean | SD | Range | Clinical Diagnostic Level % | n |
|---|---|---|---|---|---|
| **Content Moderator ($n = 19$)** | | | | | |
| PTSD Full (PDS-5) | 16.11 | 16.92 | 0 - 46 | 26.3 | 5 |
| PTSD (PDS-5)[1] | 17.00 | 16.95 | 0 - 46 | 27.8 | 5 |
| PTSD Full (PSS-I-5) | 4.86 | 4.81 | 0 - 16 | 5.3 | 1 |
| PTSD (PSS-I-5)[2] | 5.61 | 5.16 | 0 - 16 | 5.3 | 1 |
| Depression (PHQ-9) | 8.42 | 6.42 | 0 - 21 | 42.1 | 8 |
| Depression (QIDS-I) | 6.00 | 4.20 | 0 - 15 | - | - |
| Somatic Symptoms (SSS-8) | 12.84 | 5.09 | 4 - 23 | 89.5 | 17 |
| Alcohol Use (AUDIT-10) | 3.05 | 3.61 | 0 - 14 | 10.5 | 2 |
| Cannabis Use (CUDIT-R) | 3.16 | 5.03 | 0 - 17 | 15.8 | 3 |
| **Tech Support ($n = 26$)** | | | | | |
| PTSD Full (PDS-5) | 10.42 | 14.21 | 0 - 50 | 15.4 | 5 |
| PTSD (PDS-5)[3] | 15.94 | 14.90 | 0 - 50 | 22.2 | 4 |
| PTSD Full (PSS-I-5) | 0.04 | 0.20 | 0 - 1 | 0.0 | 0 |
| PTSD (PSS-I-5)[4] | 1.00 | - | - | 0.0 | 0 |
| Depression (PHQ-9) | 6.77 | 4.84 | 0 - 16 | 30.8 | 8 |
| Depression (QIDS-I) | 3.62 | 3.58 | 0 - 13 | - | - |
| Somatic Symptoms (SSS-8) | 8.54 | 5.83 | 1 - 24 | 50.0 | 13 |
| Alcohol Use (AUDIT-10) | 6.54 | 5.79 | 0 - 19 | 38.5 | 10 |
| Cannabis Use (CUDIT-R) | 2.23 | 2.87 | 0 - 17 | 7.7 | 2 |

*Note.* [1]$n$ = 18 with DSM-5 Criterion A Trauma Sample: Self-reported; [2]$n$ = 19 with DSM-5 Criterion A Trauma Sample: Interviewer; [3]$n$ = 17 with DSM-5 Criterion A Trauma Sample: Self-reported; [4]$n$ = 1 with DSM-5 Criterion A Trauma Sample: Interviewer; PDS-5: Posttraumatic Diagnostic Scale for DSM–5 ($\geq$ = 27.5, clinical cutoff); PHQ-9: Patient Health Questionnaire-9 ($\geq$ = 10, clinical cutoff); SSS-8: Somatic Symptom Scale-8 ($\geq$ = 8 moderate to severe); AUDIT-10: Alcohol Screening Questionnaire-10 ($\geq$ = 8, hazardous use cutoff); CUDIT-R: Cannabis Use Disorders Identification Test-Revised ($\geq$ = 9, clinical cutoff).

Content moderators showed slightly higher, albeit non-clinical levels, of interviewer-rated PTSD severity ($M$ = 5.61, $SD$ = 5.16) than technology support workers ($M$ = 0.04, $SD$ = .20), $F(1, 43)$ = 30.55, $p < .001$, $d$ = 1.50. Only one (5.3%) of the content moderator workers met interviewer-rated diagnostic criteria for current PTSD (PSS-I-5), whereas none of the tech support workers (0.0%) met criteria ($RR$ = 3.86). For lifetime diagnosis of PTSD based on the SCID-5, two (11.0%) of the content moderators met criteria versus none of the tech support workers (0.0%, $RR$ = 6.14).

At three month follow-up, content moderators reported more work-related intrusive memories in the past two weeks ($M$ = 11.39, $SD$ = 12.93) than tech support workers ($M$ = 4.05, $SD$ = 4.07), $F(1, 37)$ = 6.10, $p$ = .018, $d$ = 0.77) but did not differ in distress related to the worst intrusion. A similar pattern was present for PTSD severity, with no difference between groups on self-reported PTSD severity but again differences for interviewer-rated PTSD severity (Content: $M$ = 4.28, $SD$ = 3.95, Tech: $M$ = 0.58, $SD$ = 1.58; $F(1, 40)$ = 17.45, $p < .001$, $d$ = 1.21), with one content moderator and no tech support workers meeting diagnostic criteria ($RR$ = 3.94).

For self-reported depression severity (PHQ-9), content moderators and technology support workers did not differ on severity or likelihood of exceeding a clinical cut score. For interviewer-rated depression (QIDS-I), content moderators showed higher, albeit non-clinical levels of depression ($M$ = 6.00, $SD$ = 4.20) than technology support workers ($M$ = 3.62, $SD$ = 3.58), $F(1, 43)$ = 4.21, $p$ = .046, $d$ = 0.61. For structured clinical interview-rated current mood disorder (SCID-5), 15.8% ($n$ = 3) of content moderators met diagnostic criteria compared with 0.0% ($n$ = 0) of tech support workers, $\chi^2(N = 45)$ = 4.40, $p$ = .04, $RR$ = 8.22.

**Table 5.** *Work-Related Experiences, Individual Difference, and Stress Predictors of PTSD Severity (PDS-5): U.S. Sample (N = 45)*

| | | | | 95% CI | | |
|---|---|---|---|---|---|---|
| **Predictors of PTSD Severity**[1] | **B** | *SE* | **Beta** | *LB* | *UB* | *p* |
| Work-Related Experiences[2] | | | | | | |
| Hours/day Distressing Content | 0.32 | 0.96 | .06 | -1.66 | 2.28 | .74 |
| Number People Disclose Job | -0.83 | 0.64 | -.22 | -2.15 | 0.48 | .21 |
| Work-related Intrusions | 0.28 | 0.57 | .12 | -0.89 | 1.44 | .63 |
| Intrusion-related Distress | 3.01 | 1.84 | .35 | -0.76 | 6.78 | .11 |
| Work Culture | -0.14 | 0.24 | -.11 | -0.64 | 0.36 | .56 |
| Trait Individual Differences[3] | | | | | | |
| Use of Imagery (SUIS) | 0.26 | 0.32 | .14 | -0.40 | 0.92 | .43 |
| Interpersonal Reactivity (IRI) | 0.36 | 0.22 | .28 | -0.08 | 0.80 | .11 |
| Automatic Thoughts (ATQ) | 0.25 | 0.10 | .39 | 0.05 | 0.45 | .02* |
| Stress Coping and Support[4] | | | | | | |
| Daily Perceived Stress (PSS) | 0.74 | 0.38 | .37 | -0.04 | 1.52 | .06 |
| Social Support (SCS) | -0.26 | 0.19 | -.19 | -0.65 | 0.14 | .19 |
| Daily Spiritual Exp (DSES) | 0.51 | 0.19 | .37 | 0.13 | 0.89 | .01* |
| Avoidant Coping (BCOPE) | 0.50 | 0.62 | .15 | -0.76 | 1.77 | .42 |

*Note.* IRI: Interpersonal Reactivity Index; ATQ: Automatic Thoughts Questionnaire; PSS: Perceived Stress Scale; SCS: Social Connectedness Scale; DSES: Daily Spiritual Experiences; BCOPE: Brief COPE Scale.
[1] DSM-5 Criterion A trauma-exposed sample based on PDS-5 ($n$ = 35)
[2] $F(5, 27)$ = 2.15, $p$ = .09, $R^2$ = .28
[3] $F(3, 30)$ = 3.85, $p$ = .02, $R^2$ = .28
[4] $F(4, 30)$ = 6.61, $p < .001$, $R^2$ = .47

**Table 6.** *Work-Related Experiences, Individual Difference, and Stress Predictors of Depression Severity (PHQ-9): U.S. Sample (N = 45)*

| | | | | 95% CI | | |
|---|---|---|---|---|---|---|
| **Predictors of Depression Severity** | **B** | *SE* | **Beta** | *LB* | *UB* | *p* |
| Work-Related Experiences[1] | | | | | | |
| Hours/day Distressing Content | 0.32 | 0.27 | .17 | -0.23 | 0.86 | .25 |
| Number People Disclose Job | 0.01 | 0.17 | .01 | -0.33 | 0.35 | .96 |
| Work-related Intrusions | 0.11 | 0.08 | .21 | -0.06 | 0.27 | .21 |
| Intrusion-related Distress | 0.84 | 0.46 | .30 | -0.10 | 1.77 | .08 |
| Work Culture | -0.13 | 0.06 | -.32 | -0.24 | -0.01 | .03* |
| Trait Individual Differences[2] | | | | | | |
| Use of Imagery (SUIS) | 0.00 | 0.08 | .00 | -0.15 | 0.16 | .98 |
| Interpersonal Reactivity (IRI) | 0.12 | 0.05 | .25 | 0.01 | 0.22 | .03* |
| Automatic Thoughts (ATQ) | 0.17 | 0.03 | .72 | 0.12 | 0.22 | <.001* |
| Stress Coping and Support[3] | | | | | | |
| Daily Perceived Stress (PSS) | 0.40 | 0.09 | .55 | 0.22 | 0.58 | <.001* |
| Social Support (SCS) | 0.02 | 0.05 | .04 | -0.08 | 0.11 | .70 |
| Daily Spiritual Exp (DSES) | 0.04 | 0.03 | .11 | -0.03 | 0.10 | .26 |
| Avoidant Coping (BCOPE) | 0.38 | 0.14 | .34 | 0.10 | 0.66 | .009* |

*Note.* IRI: Interpersonal Reactivity Index; ATQ: Automatic Thoughts Questionnaire; PSS: Perceived Stress Scale; SCS: Social Connectedness Scale; DSES: Daily Spiritual Experiences; BCOPE: Brief COPE Scale.
[1] $F(5, 35)$ = 4.01, $p$ = .006, $R^2$ = .36
[2] $F(3, 40)$ = 17.02, $p < .001$, $R^2$ = .56
[3] $F(4, 40)$ = 20.34, $p < .001$, $R^2$ = .67

For the presence of major depressive disorder (SCID-5), while there was no difference in current MDD diagnosis, content moderators were more likely to have a past diagnosis of MDD (57.9%, $n = 11$) than tech support workers (26.9%, $n = 7$), $\chi^2(N = 46) = 4.39, p = .036, RR = 2.15$. At three month follow-up, while the pattern was similar to baseline, there were no differences between content moderators and tech support workers on depression severity (PHQ-9, QIDS-I) or mood disorder or MDD diagnoses (SCID-5).

**Somatic Symptoms, Alcohol Use, and Cannabis Use.** For somatization (SSS-8), content moderator workers reported higher somatic symptoms ($M = 12.84, SD = 5.09$) than tech support workers ($M = 8.54, SD = 5.83$), $F(1, 43) = 6.64, p = .013, d = 0.79$. Further, content moderators were also more likely to report moderate or higher levels of somatic symptoms (89.5%, $n = 17$) than tech support workers (50.0%, $n = 13$), $\chi^2(N = 45) = 7.70, p = .006, RR = 1.42$. At three month follow-up, while the same pattern was present, the difference no longer approached significance.

In contrast, tech support workers reported higher alcohol use problems (AUDIT-10: $M = 6.54, SD = 5.79$) than content moderators ($M = 3.05, SD = 3.61$), $F(1, 43) = 5.34, p = .026, d = 0.72$. Further, tech support workers were also more likely to report hazardous alcohol use (38.5%, $n = 10$) than content moderators (10.5%, $n = 2$), $\chi^2(N = 45) = 4.38, p = .036, RR = .34$. For clinical interviewer rated presence of an alcohol use disorder (SCID-5), both lifetime (tech support: 42.3% [$n = 11$], content moderators: 26.3% [$n = 5$], $RR = .70$) and current diagnoses (tech support: 15.4% [$n = 4$], content moderators: 5.3% [$n = 1, RR = .38$]) mirrored this, but were not statistically different. There were no differences on cannabis use severity (CUDIT-R) clinical cut-off defined cannabis use disorder. Differences on clinical interviewer-rated (SCID-5) presence of lifetime or current cannabis use disorder were not observed. At three month follow-up, on problematic alcohol use, there was no difference between content moderators and tech support workers, though content moderators were more likely to report problematic cannabis use (content moderator: $M = 3.72, SD = 5.39$, tech support: $M = 1.00, SD = 2.43$; $F(1, 38) = 4.68, p = .037, d = 0.65$). Further, content moderators were also more likely to report moderate to severe cannabis use (16.7%, $n = 3$) than tech support workers (0.0%, $n = 0$), $\chi^2(N = 41) = 4.14, p = .042, RR = 8.22$.

### Predictors of Mental Health
**Work Experience Predictors.** When examining work-related experiences as predictors of mental health, as seen in Table 5, for those who have experienced a DSM-5 Criterion A trauma, there were no predictors associated with higher PTSD severity (PDS-5). However, as seen in Table 6, worse workplace culture ($b = -.32, p = .03$) was associated with higher depression severity (PHQ-9). Baseline work-related experience variables did not predict PTSD (PDS-5) or depression severity (PHQ-9) three months later.

**Individual Difference Trait Predictors.** When examining trait individual difference predictors of mental health, as seen in Table 5, for those who have experienced a DSM-5 Criterion A trauma, higher negative automatic thoughts ($b = .39, p = .02$) was associated with higher PTSD severity (PDS-5). As seen in Table 6, both higher interpersonal reactivity/empathy ($b = .25, p = .03$) and higher negative automatic thoughts ($b = .72, p < .001$) were associated with higher depression severity (PHQ-9).

Higher baseline negative automatic thoughts predicted both higher PTSD severity (PDS-5; $b = .64, p < .001, F(3, 20) = 8.05, p = .001, R^2 = .55$) and higher depression severity (PHQ-9; $b = .68, p < .001, F(3, 35) = 12.93, p < .001, R^2 = .53$) three months later.

**Stress, Social Support, and Coping Predictors.** When examining stress, support, and coping-related predictors of mental health, as seen in Table 5, for those who had experienced a DSM-5 Criterion A trauma, higher daily spiritual experiences ($b = .37, p = .01$) were associated with higher PTSD severity (PDS-5). As seen in Table 6, higher perceived stress in the last month ($b = .55, p < .001$) and higher avoidant coping ($b = .34, p = .009$) were associated with higher depression severity (PHQ-9).

Higher baseline avoidant coping (BCOPE; $b = .50, p = .02$) predicted higher PTSD severity (PDS-5) three months later, $F(4, 20) = 5.62, p = .003, R^2 = .53$. Similarly, higher baseline previous month stress (PSS; $b = .41, p = .02$) and higher avoidant coping (BCOPE; $b = .34, p = .04$) predicted depressive severity three months later, $F(4, 35) = 8.96, p < .001, R^2 = .51$.

## DISCUSSION

Humans will remain essential arbiters of appropriateness in digital environments even as automated systems improve, whether for social media, search, or adjacent "data enrichment" work such as AI red-teaming. In two samples, an international contractor sample and a U.S. sample with a comparison group and clinical interviews, content moderators showed elevated probable depression, PTSD, and marked somatic symptom burden. In the U.S. sample, moderators had higher interviewer-rated PTSD severity, were eight times more likely to meet diagnostic criteria for a current mood disorder, and over two times more likely to report a lifetime MDD diagnosis than the comparison group. Crucially, the main risks were not limited to exposure "dose." Notably, the number of hours per day spent reviewing distressing content was not a robust predictor in these samples, likely due to restricted exposure ranges reflecting employer caps. In contrast, intrusion-related distress, negative automatic thoughts, higher perceived stress, avoidant coping, and, for depression, poorer perceived workplace culture were consistent predictors, including prospectively at three months. These results converge with occupational health models in which work design (e.g., demand, control, support) and cognitive-affective processes co-determine risk, and they direct our focus to organizational and structural levers of change. By pairing a large international CM cohort with a U.S. cohort that includes gold-standard diagnostic interviews, we show a potential link with clinical levels of mental health problems and associated diagnostic categories relevant to care and benefits. Further, we identify modifiable targets such as organizational culture, pacing and rotation practices that shape intrusive memories, and negative beliefs, stress, and avoidant coping that are amenable to brief evidence-based interventions.

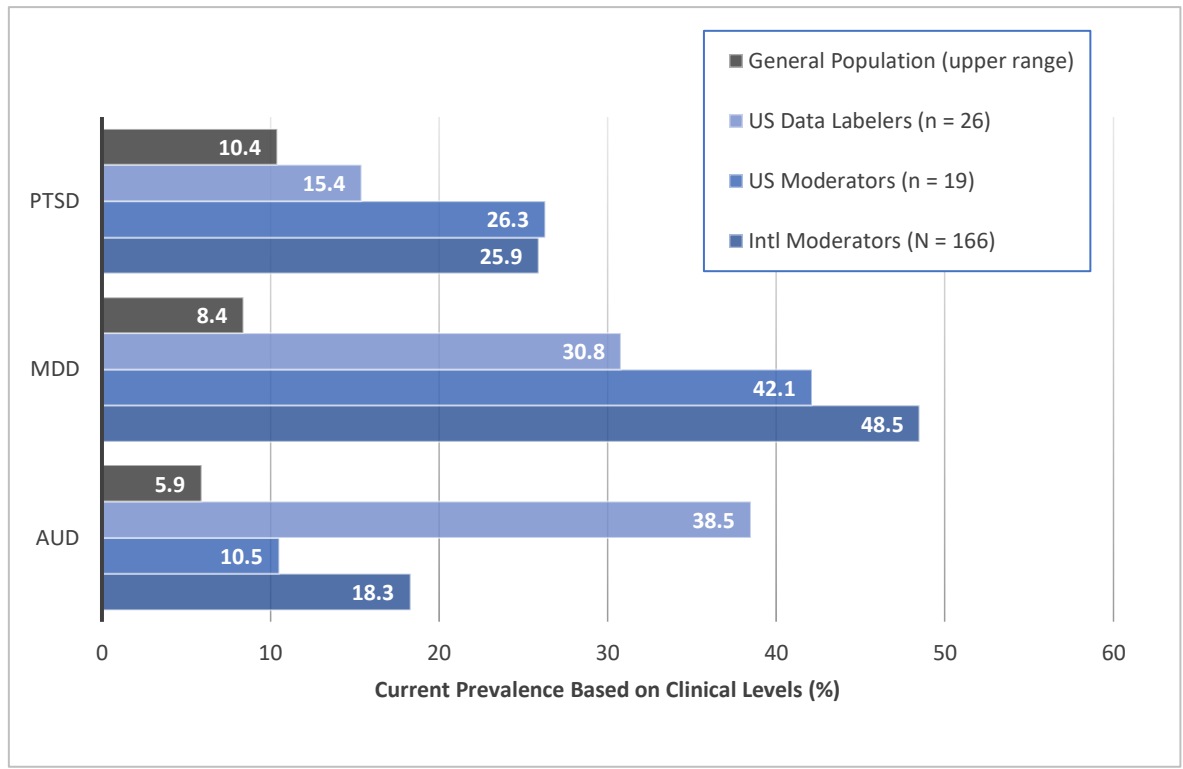

**Figure 1.** *Percentage of Individuals with Clinical Levels of PTSD, Major Depression (MDD), and Alcohol Use Disorder (AUD) Across Samples*

The rates of clinical levels of psychopathology observed across both samples were elevated relative to the general population. In particular, elevated rates of clinical-level depression in both the international (48.5%) and U.S. samples (42.1%), and in clinical interviewer diagnosed major depression (15.8% [content moderators] % vs 0% [tech support]), exceed rates observed in global samples (e.g., 2.2% - 10.4% past 12-month prevalence; Kessler & Bromet, 2013). This pattern of elevation, though to a lesser magnitude, was also observed in clinical levels of posttraumatic stress disorder (25.9% [international], 26.5% [U.S. based]) and clinical levels of alcohol use (18.3% [international], 10.5% [U.S. based]) compared to the general population (0.2% - 8.4% past 12-month PTSD; 0.1% - 5.9% past 12-month alcohol use disorder; Glantz et al., 2020; Koenen et al., 2017). See Figure 1. Elevated rates of depression are potentially most concerning and replicated in the clinical interviewer data, with U.S.-based content moderators being 8.2 times more likely to report a current mood disorder than data labelers/technology support workers and over half (57.9%) of them reporting a lifetime history of MDD. These elevated rates are consistent with a brief, cross-sectional survey that also showed elevated, more global psychological distress, a dose-dependent association between frequency of exposure to distressing content and psychological distress, and qualitative analyses suggesting similar concerns (Spence et al., 2023; 2024). It should be noted that this prior research and the present study are not causal and do not directly link content moderation work to elevated mental health problems. That said, despite growing awareness of the need to support the wellness of content moderators, existing efforts within corporations may be inadequate to counter the potentially detrimental mental health effects of this work for some workers.

Poorer workplace culture, work-related intrusion distress, and perceived stress also emerged as predictors of PTSD and depression. Indeed, when looking at the relationship between workplace environment and depressive symptoms, job strain, typically defined as high psychological demand and low decision making latitude, consistently emerges as one of the strongest predictors of depressive symptoms (e.g., Theorell et al., 2015). Work environment can potentially be enhanced by a range of strategies, including paying well, creating supportive teams, and leadership training in recognizing and reducing workplace stress (Goetzel et al., 2018; Wu et al. 2021). This also likely includes a mental-health informed work culture, integrating preventative interventions, such as what is known about reducing stress (e.g., relaxation, mindfulness), addressing unhelpful negative cognitions (e.g., cognitive restructuring), and reducing avoidant coping (e.g., harm reduction for substance use). Further, distress related to work-related intrusions can potentially be reduced by altering content moderation procedures, such as systematically examining factors like moderation load amounts, timing within a day or a week, massing/grouping of content, or altering playback/moderation format and images, and monitoring fatigue, among others (e.g., Li et al., 2025). Technological systems that control moderation environment have considerable promise in potentially reducing the development of intrusive memories.

The literature generally infers a dose-response relationship between traumatic content exposure and psychological harm, though this relationship is complex. Our findings are consistent with a conditional dose effect: when exposure range is limited (e.g., 2–4 hours/day), organizational context (team support, supervisor practices, autonomy) and intrusion-related distress may dominate as proximal drivers. This underscores an intervention strategy that couples exposure management with work design improvements. Content-related distress and fatigue could be mitigated by carefully monitoring and limiting time spent on content moderation within a workday and week. Both employers and moderators themselves need to know when in a particular shift they need to stop and shift to other work for longer periods of time (e.g., weeks, months). Notably, there was only a small, non-significant association for hours monitoring distress-related content per day, arguing that other factors besides "dose" of distressing content may be more important. However, it should be noted that there was little variability across samples, as content moderators in this study typically reported working around four hours or less per day on distressing content (mode = 2hrs/day international sample, 3 hrs/day U.S. sample) limiting the range to detect a potential relationship. Indeed, this may reflect employers strategically limiting the amount of content moderation per day but also potentially argues that a stronger relationship could emerge as content moderation time increases. There is accumulating evidence that the amount of work-related trauma exposure, both direct or more indirect exposure, is associated with an increased likelihood of PTSD (Coenen & van der Molen, 2021). Mitigating the amount or impact of this exposure could reduce the number of depression and PTSD cases within specific higher-risk work populations.

Across both samples, one of the most consistent predictors of PTSD and depression was higher perceived ongoing life stress. Indeed, both samples showed heightened clinical elevations of somatic symptoms (e.g., headaches, stomach problems, low energy), which are physical symptoms commonly associated with stress. Life stress likely causally contributes to risk for depression (e.g., Bjørndal et al., 2023), and it may set up content moderators to utilize unhelpful coping with work-related stress. Work-related stress and fatigue, signaling the need for a potential break or shift of work content, can potentially be addressed by examining proxies for stress during content moderation itself, such as slower reaction times, reduced efficiency, or even asking for check-in distress ratings from the moderator, building in confidentiality/privacy barriers that would allow feedback to the individual but would make specific linkage to the individual difficult and not shared with the employer. More broadly, routine monitoring of work-related stress and monitoring of mental health wellness may help individuals identify when longer breaks in content moderation work are needed. Relatively robust and low-burden stress reduction interventions already exist, including progressive muscle relaxation (e.g., Chellew et al., 2015) and mindfulness-based approaches (e.g., Janssen et al., 2018). Further, creating working conditions that buffer against work-related stressors, such as opportunities to connect interpersonally with coworkers, may be beneficial. In the present samples, better perceived social support and the higher disclosure of difficult aspects of their job with others were often also associated with lower PTSD and depression symptoms.

Negative beliefs emerged, by far, as the strongest and most consistent predictor of higher PTSD and depression severity, across both samples cross-sectionally and in predicting the presence of PTSD and depression three months later. The persistent presence of these negative beliefs (e.g., "I feel like I am up against the world," "No one understand me," "I'm a loser"), often reflecting negative interpretation biases or catastrophizing (e.g., Nieto et al., 2020), may be an early warning signal for content moderators themselves and those in their wellness support networks to pay attention to and intervene. This linkage across negative beliefs, subsequent psychopathology, and examining and restructuring these beliefs to reduce depression is well established (e.g., Everaert et al., 2017; Diehle et al., 2014). Training in brief cognitive restructuring, addressing rumination, or behavioral activation interventions may help prevent the development of PTSD and depression (e.g., Bedard-Gilligan et al., 2022; Cuijpers & Reynolds, 2022; Topper et al., 2010). Another potential early warning signal may be the more frequent use of avoidant coping strategies to manage stress, which emerged as a consistent and relatively strong predictor of both PTSD and depression severity across samples. This coping style includes engaging in distracting activities (e.g., playing video games, using substances), discounting the perceived stress, and avoiding difficult situations in order to prevent negative thoughts and emotions. In first responders, these strategies similarly are related to worse mental health and job burnout, while approach-related, active coping strategies such as positive reinterpretations and emotional support have emerged as protective factors (Díaz-Tamayo et al., 2022). Further, training in harm reduction and use of motivational interviewing techniques may help prevent the development of substance use disorders (e.g., Larimer et al., 2022; Mattoo et al., 2018).

One of the strengths of this study was the inclusion of both an international, contractor-based content moderator sample and a U.S.-based content moderator sample with a technology support/data labeling control condition. Further, we used well established, psychometrically validated measures of psychopathology and, in the U.S. sample, conducted gold-standard clinical diagnostic interviews using well-trained mental health assessors. The convergence across the samples and replication with clinical diagnostic interviews strengthens both the quality and generalizability of the findings. While the control condition did not differ from the content moderators in income, education, or employment with a major technology company, the nature of their work may not be an ideal comparator. Specifically, customer support workers may have pre-existing high levels of psychological distress (Doellgast & O'Brady, 2020; Raja & Bhasin, 2014; Shih et al., 2014), deal with frustrating client problems daily, and often work in a less satisfactory environment with stricter work schedules (Gonçalves-Candeias et al., 2021). Of note, some of this data was collected during later phases of the COVID-19 pandemic with many of these workers working at home (data collection: June 2021-August 2024). Factors such as higher social isolation, boredom, anger, and confusion may have contributed to heightened stress and psychopathology (Brooks et al., 2020; Tavares, 2017). However, the rates observed in the present studies are elevated even in comparison to those specifically focused on the impacts of the early phases of COVID and healthcare workers (e.g., Dragioti et al., 2022). Lastly, neither of the samples assessed workers prior to the start of their content moderation work and thus cannot rule out pre-existing characteristics driving the present results. The pattern across both samples and the ability of key individual difference factors at baseline, namely, ongoing stress, negative beliefs, and avoidant coping predicting PTSD and depression three months later, add to the confidence that these are important factors for targeted intervention at both the individual and structural levels.

The present study documents potential mental health impacts for those who perform content moderation. Much of this work is carried out as pay-per-task, internationally out-sourced, and/or individual work (conducted at home, computer cafe, etc.). Care should be given both for those content moderator workers working within technology companies and those working alone or with contractors to support mental health. This research does not say that being a content moderator causes mental health problems, but it does suggest that it potentially puts a subset of workers at heightened risk for this impact. There are potential implications beyond content moderation, including AI red-teaming and data enrichment work. Red-teaming exposes workers to potentially harmful or disturbing model outputs in adversarial testing contexts. Our results suggest that a similar risk constellation of repetitive exposure plus organizational pressures and inadequate supports could emerge in red-teaming without proactive design. This suggests exporting the organizational and interface-level protections proposed here to human-in-the-loop content work, including red-teaming.

Parameters and support around this work need to be better understood and best practices determined and implemented. To do this, we call for: 1) intervention trials testing organizational and interface changes (e.g., adaptive task pacing; visual/audio design modifications to attenuate sensory vividness and reduce intrusive memory encoding; confidential distress dashboards with individual feedback loops); 2) quasi-experimental designs to estimate impacts of policy changes (e.g., workload or exposure limits); 3) longitudinal research following workers from onboarding through exit to identify cumulative effects of digital work environments; and (4) predictive modeling of hours-increase scenarios under varying culture and support conditions to forecast risk and guide organizational policy. Finally, we advocate for cross-sector collaboration to design high-performance "red-teaming" environments that avoid replicating avoidable harms and innovate safely.

## ACKNOWLEDGMENTS

This research was supported by Microsoft Corporation, Inc. Microsoft had no role in study design, data analysis, or research dissemination. We would like to acknowledge Gustavo Basualdo and Mary Gray, Ph.D. for their ongoing feedback over the years of conducting this research and, later, comments on the final manuscript. Correspondence concerning this study should be addressed to Lori Zoellner, Department of Psychology, University of Washington, Box 351525, Seattle, WA, 98195. Email: zoellner@uw.edu

**Supplemental Materials 1**
**Thoughts and Beliefs Related to Work Environment**
Read the following statements and provide a rating of the extent to which you agree that these statements describe your work environment DURING THE PAST MONTH.

|  | Strongly Disagree 1 | 2 | 3 | Neither Agree nor Disagree 4 | 5 | 6 | Strongly Agree 7 |
|---|---|---|---|---|---|---|---|
| 1. I am happy with my work environment. | ○ | ○ | ○ | ○ | ○ | ○ | ○ |
| 2. I feel like I am a member of a strong team. | ○ | ○ | ○ | ○ | ○ | ○ | ○ |
| 3. I feel support from my co-workers. | ○ | ○ | ○ | ○ | ○ | ○ | ○ |
| 4. I have access to mental health resources at work (e.g., wellness coach, therapy services). | ○ | ○ | ○ | ○ | ○ | ○ | ○ |
| 5. Wellness is part of my team's culture. | ○ | ○ | ○ | ○ | ○ | ○ | ○ |
| 6. Wellness is part of my company's culture. | ○ | ○ | ○ | ○ | ○ | ○ | ○ |
| 7. It is important to me to engage in good wellness practices (e.g., self-monitoring journal). | ○ | ○ | ○ | ○ | ○ | ○ | ○ |
| 8. My work has a positive impact on the world. | ○ | ○ | ○ | ○ | ○ | ○ | ○ |
| 9. My work makes online a safer place. | ○ | ○ | ○ | ○ | ○ | ○ | ○ |
| 10. My work makes me feel that the world is unjust.* | ○ | ○ | ○ | ○ | ○ | ○ | ○ |
| 11. My work makes me feel that there are evil people in the world.* | ○ | ○ | ○ | ○ | ○ | ○ | ○ |
| 12. My work is dangerous to my health.* | ○ | ○ | ○ | ○ | ○ | ○ | ○ |
| 13. My supervisor understands what it's like to perform my work. | ○ | ○ | ○ | ○ | ○ | ○ | ○ |

---

*Items #10, #11, #12 are reverse coded.
Gauthier, G. M., Ali, E., Asim, A., Cornell-Maier, S., & Zoellner, Lori A. (2025). I've seen enough: Measuring the toll of content moderation on mental health. e-Print arXiv.org